\documentclass[aps,pra,twocolumn,superscriptaddress]{revtex4-1}
\usepackage{graphicx}
\usepackage{dcolumn}
\usepackage{bm}
\usepackage[english]{babel}
\usepackage[utf8]{inputenc}
\usepackage[T1]{fontenc}       	
\usepackage{mathptmx}
\usepackage{upgreek}
\usepackage{float}
\begin{document}

\preprint{AIP/123-QED}

\title{Spin Measurements of NV Centers Coupled to a Photonic Crystal Cavity}

\author{T. Jung}
\author{J. Görlitz}
\author{B. Kambs}
\affiliation{
Universit\"at des Saarlandes, Fachrichtung Physik, Campus E\,2.6, 66123 Saarbr\"ucken, Germany.
}
\author{C. Pauly}
\affiliation{
Universit\"at des Saarlandes, Fachrichtung Materialwissenschaft und Werkstofftechnik, Campus D\,3.3, 66123 Saarbr\"ucken, Germany.
}
\author{N. Raatz}
\affiliation{
Universit\"at Leipzig, Angewandte Quantensysteme, Linn\'{e}straße 5, 04103 Leipzig, Germany.
}
\author{R. Nelz}
\author{E. Neu}
\affiliation{
Universit\"at des Saarlandes, Fachrichtung Physik, Campus E\,2.6, 66123 Saarbr\"ucken, Germany.
}
\author{A. M. Edmonds}
\author{M. Markham}
\affiliation{
Element Six Global Innovation Centre, Fermi Avenue, Harwell Oxford, Didcot, Oxfordshire OX11 0QR, United Kingdom.
}
\author{F. Mücklich}
\affiliation{
Universit\"at des Saarlandes, Fachrichtung Materialwissenschaft und Werkstofftechnik, Campus D\,3.3, 66123 Saarbr\"ucken, Germany.
}
\author{J. Meijer}
\affiliation{
Universit\"at Leipzig, Angewandte Quantensysteme, Linn\'{e}straße 5, 04103 Leipzig, Germany.
}
\author{C. Becher}
 \email{christoph.becher@physik.uni-saarland.de}
\affiliation{
Universit\"at des Saarlandes, Fachrichtung Physik, Campus E\,2.6, 66123 Saarbr\"ucken, Germany.
}

\date{\today}

\begin{abstract}
Nitrogen-vacancy (NV) centers feature outstanding properties like a spin coherence time of up to one second as well as a level structure offering the possibility to initialize, coherently manipulate and optically read-out the spin degree of freedom of the ground state. However, only about three percent of their photon emission are channeled into the zero phonon line (ZPL), limiting both the rate of indistinguishable single photons and the signal-to-noise ratio (SNR) of coherent spin-photon interfaces. We here report on the enhancement of the SNR of the optical spin read-out achieved by tuning the mode of a two-dimensional photonic crystal (PhC) cavity into resonance with the NV-ZPL. PhC cavities are fabricated by focused ion beam (FIB) milling in thin reactive ion (RIE) etched ultrapure single crystal diamond membranes featuring modes with $Q$-factors of up to $8250$ at mode volumes below one cubic wavelength. NV centers are produced in the cavities in a controlled fashion by a high resolution atomic force microscope (AFM) implantation technique. On cavity resonance we observe a lifetime shortening from $9.0\,\mathrm{ns}$ to $8.0\,\mathrm{ns}$ as well as an enhancement of the ZPL emission by almost one order of magnitude. Although on resonance the collection efficiency of ZPL photons and the spin-dependent fluorescence contrast are reduced, the SNR of the optical spin read-out is almost tripled for the cavity-coupled NV centers.
\end{abstract}

\maketitle

\section{\label{sec:Intro}Introduction}

The nitrogen-vacancy (NV) center, a point defect in diamond consisting of a lattice vacancy and an adjacent nitrogen substitution, has attracted a lot of interest during the past years owing to its outstanding optical and spin properties.\cite{Doherty2013} The triplet ground state exhibits two sublevels attributed to two spin projections $m_s=0$ and $m_s=\pm1$ of the NV electron spin.\cite{Doherty2011} In addition to an ultralong spin coherence time of more than one second at liquid helium temperatures,\cite{Bar-Gill2013,Abobeih2018} the NV center features spin-conserving optical transitions.\cite{Robledo2011} Furthermore, the electron spin may be coherently manipulated by microwave signals \cite{Jelezko2004a} and purely optically initialized as well as read-out.\cite{Gruber1997} Spin initialization and read-out are enabled by a spin-selective intersystem crossing (ISC) towards the singlet system: the long lifetime in the singlet system facilitates a spin-dependent fluorescence and a preferred decay towards the $m_s=0$ ground state,\cite{Kalb2018,Thiering2018} allowing for fast spin initialization. Spin polarizations of about $80\,\%$ at room temperature \cite{Neumann2010,Robledo2011a} and over $99\,\%$ at liquid helium temperature \cite{Robledo2011} may be reached, as well as spin-dependent fluorescence contrasts of up to $30\,\%$ for an optical spin read-out under non-resonant laser excitation.\cite{Dreau2011}\\
Besides the spin-dependent fluorescence contrast, the reliability of an optical spin read-out, designated as the signal-to-noise ratio (SNR), also depends on the detected photon count rate. The SNR for an optical measurement distinguishing between the two possible spin projections $m_s=0$ and $m_s=\pm 1$ of a NV center is defined as\cite{Steiner2010}
\begin{eqnarray}
	\mathrm{SNR} & = & \frac{N_0-N_1}{\sqrt{N_0+N_1}} \ .
	\label{eq:SNRDefinition}
\end{eqnarray}
Here $N_0$ ($N_1$) is the expectation value for the detected photon count rate when preparing the NV center in the spin projection $m_s=0$ ($m_s=\pm1$). Due to the Poisson-distribution of the photon count rate we expect that the random variables $N_0$ and $N_1$ are also Poisson-distributed,\cite{Wolf2015} hence their difference is Skellam-distributed with variance $\sigma^2=N_0+N_1$.\cite{Skellam1946} As apparent in equation (\ref{eq:SNRDefinition}), the SNR is mainly determined by the difference $N_0-N_1$. Normalized by $N_0$, we obtain the already mentioned spin-dependent fluorescence contrast $C=(N_0-N_1)/N_0$. With this measure we get the relation
\begin{eqnarray}
	\mathrm{SNR} & = & \sqrt{N_0}\cdot \frac{C}{\sqrt{2-C}} \ .
	\label{eq:SNRmitKontrast}
\end{eqnarray}
As visible from equation (\ref{eq:SNRmitKontrast}), a large SNR of optical spin read-out requires both a large collected photon rate and a large contrast of spin-dependent fluorescence.\cite{Hopper2018} In view of applications, a sufficiently large SNR facilitates the use of NV centers as quantum sensors for temperature,\cite{Kucsko2013} pressure \cite{Doherty2014} as well as magnetic fields on the nanoscale \cite{Balasubramanian2008,Fuchs2008,Maze2008,Bernardi2017} or the verification of fundamental principles of quantum mechanics as for example demonstrated by a loophole free Bell test.\cite{Hensen2015} Furthermore, a high enough SNR in combination with the long spin coherence time enables the coupling of NV centers to nearby nuclear spins used as quantum bits \cite{Neumann2010a,Maurer2012,Reiserer2016,Yang2016} or as building blocks of quantum repeaters.\cite{Nemoto2016,Rozpedek2019} In particular, for an aspired scaling of quantum systems towards quantum networks,\cite{Wehner2018} a speed-up of the entanglement generation is a central requirement and hence a preferably large SNR desirable.\\
The photon detection rate is usually limited by a non-perfect photon collection. Consequently, several approaches to enhance the collection efficiency have been followed by modifying the directivity of emission such as using an optimal crystal orientation,\cite{Jamali2014} fabricating solid immersion lenses around NV centers \cite{Hadden2010,Robledo2011} or incorporating NV centers in nanopillars,\cite{Maletinsky2012,Appel2016} nanowires,\cite{Babinec2010} waveguides \cite{Hausmann2012,Mouradian2015} or metalenses.\cite{Grote2017} Furthermore, the collection efficiency can be modified by coupling NV centers to whispering gallery \cite{Faraon2011,Hausmann2012} or photonic crystal (PhC) cavities\cite{Faraon2012,Hausmann2013,Li2015,Riedrich-Moeller2015,Schroeder2017} as well as to plasmonic structures.\cite{Choy2011,Bogdanov2017} In addition, by such a coupling the local density of states at the emitter's position and hence its spontaneous emission rate may be enhanced or, correspondingly, the spontaneous emission lifetime reduced by the Purcell-factor $F$.\cite{Purcell1946} In addition to a reduced lifetime, the coupling of NV centers to a cavity has the advantage that more than the usual $3\,\%$ of the photons are emitted into the zero phonon line (ZPL). Hence, entanglement generation by interference of ZPL photons at a beam splitter \cite{Bernien2012,Sipahigil2012,Bernien2013,Hensen2015} may be further sped up by coupling to a cavity featuring a mode in resonance with the NV-ZPL. However, a modification of the population dynamics, such as Purcell enhancement of emission, also influences the spin-dependent fluorescence contrast $C$. Bogdanov et al. showed for NV ensembles in nanodiamonds, that the contrast decreases when reducing the emitter’s lifetime by coupling to plasmonic islands.\cite{Bogdanov2017} Furthermore, Babinec et al. in a theoretical study found that the spin read-out SNR achieves a maximum value for Purcell-factors on the order of $1$.\cite{Babinec2012}\\
In this article, we report on the SNR enhancement achieved by coupling NV centers to a two-dimensional PhC cavity. Starting from ultrapure single crystal diamond membranes bonded on a sacrificial silicon substrate and thinned by reactive ion etching (RIE), two-dimensional PhC cavities are fabricated by focused ion beam (FIB) milling at thoroughly characterized and carefully selected membrane positions.\cite{Jung2016} NV centers are subsequently incorporated into the PhC cavities by a high resolution AFM implantation technique.\cite{Pezzagna2010a} A FIB-milled hole in the AFM-tip serves as an aperture, which enables the accurate implantation of nitrogen into the cavities. We activate the NV centers by extensive post processing, i.\,e. annealing and cleaning procedures. We here also report on the deterministic spectral tuning of a cavity mode in resonance with the NV-ZPL using thermal oxidation of diamond as well as condensation of residual gas in the cryostat. We finally experimentally measure and theoretically simulate the change in spin-dependent fluorescence contrast $C$ and SNR on resonance.

\section{\label{sec:Nanofab} Cavity fabrication and NV incorporation}

\begin{figure}[!tb]
     \includegraphics[width=7cm]{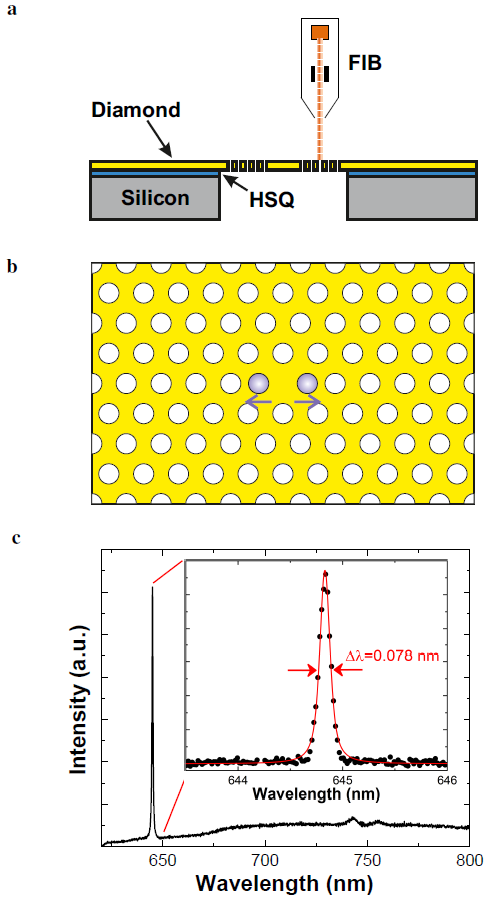}
	\caption{(a) A silicon substrate structured with windows on the scale of $150\times150\,\upmu\mathrm{m}^2$ is spin-coated with a $50\,\mathrm{nm}$ thick HSQ layer. A CVD-grown, ultrapure, single crystal diamond film is positioned on the HSQ layer generating an air suspended diamond membrane. After RIE thinning from the topside the diamond has a final thickness of a few hundred nanometers. PhC cavities are produced by FIB milling at pre-selected defect-free spots with a suitable thickness. (b) Scheme of a M0-cavity generated by a slight shift of two adjacent holes. (c) Photoluminescence (PL) spectroscopy of a M0-cavity at room-temperature. The cavity mode at $644.8\,\mathrm{nm}$ features a FWHM of $0.078\,\mathrm{nm}$, corresponding to a $Q$-factor of $8250$. The spectra were taken with an integration time of $30\,\mathrm{s}$ under non-resonant laser excitation at $532\,\mathrm{nm}$ and a laser power of $1\,\mathrm{mW}$. }
	\label{fig:Nanofabrikation}
\end{figure}

\subsection{Sample system with FIB-milled cavities}

The Purcell-factor $F$, quantifying the lifetime reduction achieved by the emitter-cavity-coupling is directly proportional to the ratio of quality factor $Q$ and modal volume $V$ of a PhC cavity mode. The $Q$-factor is strongly dependent on the precision of fabrication and deviations from the design parameters. In order to produce PhC cavities with modes featuring a high $Q$-factor and a spectral position close to the NV-ZPL, diamond membranes with a precise thickness are required. However, due to polishing wedges as well as local deviations in the etching rate during RIE-thinning the thickness typically varies over the diamond film by several hundred nanometers.\cite{Jung2016} Therefore, the thinned membranes need to be characterized carefully in order to select suitable spots for the subsequent cavity production. The challenge in FIB milling of the PhC array is the fabrication of regularly hole patterns with vertical hole sidewalls. Conical shapes with typically observed inclination angles of as large as $9^{\circ}$ may degrade the $Q$-factor by one order of magnitude.\cite{Tanaka2003,Jung2016} The fabrication process and characterization methods are described in detail in our previous publication \cite{Jung2016}.\\
The sample system is depicted in Fig.~\ref{fig:Nanofabrikation}a. As starting material we use $30\,\upmu\mathrm{m}$ thick, chemical vapor deposition (CVD) grown, (001)-oriented, ultrapure, single crystal diamond membranes (electronic grade quality, ElementSix) with a nitrogen concentration below $5\,\mathrm{ppb}$. At first, the diamond film is etched by RIE in an $\mathrm{Ar/O}_2$-plasma to remove $5\,\upmu\mathrm{m}$ of the surface material as its quality is degraded due to the polishing process.\cite{Volpe2009} After acid-cleaning, the remaining $20\,\upmu\mathrm{m}$ thick diamond film is, mediated by a spin-coated $50\,\mathrm{nm}$ thick layer of hydrogen silsesquioxane (HSQ XR-1541-002, DowCorning), bonded to a silicon substrate containing pre-fabricated windows. Curing the sample system at $600\,^{\circ}\mathrm{C}$ renders the bond persistent during the post-processing and cleaning procedure required after FIB milling as well as for removing damages in the diamond crystal lattice after nitrogen implantation (see section \ref{NVimplant}). Furthermore, the stable bond allows measurements at liquid helium temperatures. The final membrane thickness of a few hundred nanometers is finally reached by further RIE thinning from topside.\\
The membranes are subsequently characterized by laser-scanning microscopy, cross-section measurements, and quantitative dispersive X-ray spectroscopy. The combination of these methods allows us to map the thickness of the diamond films as well as the surface structure with high resolution, enabling the selection of defect-free spots featuring a suitable thickness for the fabrication of PhC cavities.\\
M0-cavities generated by a shift of two adjacent holes (Fig.~\ref{fig:Nanofabrikation}b) are fabricated by FIB milling. The hole radii of $R=68\,\mathrm{nm}$ and the lattice constant of $a=250\,\mathrm{nm}$ of the photonic crystal array are chosen such, that the cavity modes match the NV-ZPL position at $637\,\mathrm{nm}$. In addition, the M0-cavity is optimized by slight changes in position and/or radii of holes close by the point-defect.\cite{Riedrich-Moeller2010} The simulated $Q$-factor of the cavity mode is $320\,000$ at a mode volume of $0.35\,(\lambda/n)^3$.\\
Optimizations in the FIB milling process as for instance the use of overmilling and drift control programs, realizing a chamber pressure below $5\cdot10^{-6}\,\mathrm{mbar}$ and a temperature stabilization of the ion-column as well as the deposition of a metal protection layer prior to FIB milling reduce the conical hole shape to below $4^{\circ}$, optimize the hole positions within the pattern and lead to sharp and well-defined hole edges. After an extended post-processing, consisting of annealing steps in vacuum at $1000\,^{\circ}\mathrm{C}$ and acid-cleaning, the fabricated cavities are analyzed in a home build confocal setup. The PhC cavities feature modes with $Q$-factors up to $8250$, as depicted in Fig.~\ref{fig:Nanofabrikation}c. Hence, the obtained $Q$-factors reach the same order of $Q$-factors of two-dimensional PhC cavities with small modal volumes (around one cubic wavelength) and a spectral mode position close to the NV-ZPL as currently observed from RIE fabrication methods.\cite{Faraon2012,Li2015a,Schroeder2017}

\subsection{\label{NVimplant}NV incorporation by AFM implantation}

This section summarizes how we generate NV centers located at the field maximum of the PhC cavity.
\begin{figure*}[!tb]
     \includegraphics[width=16cm]{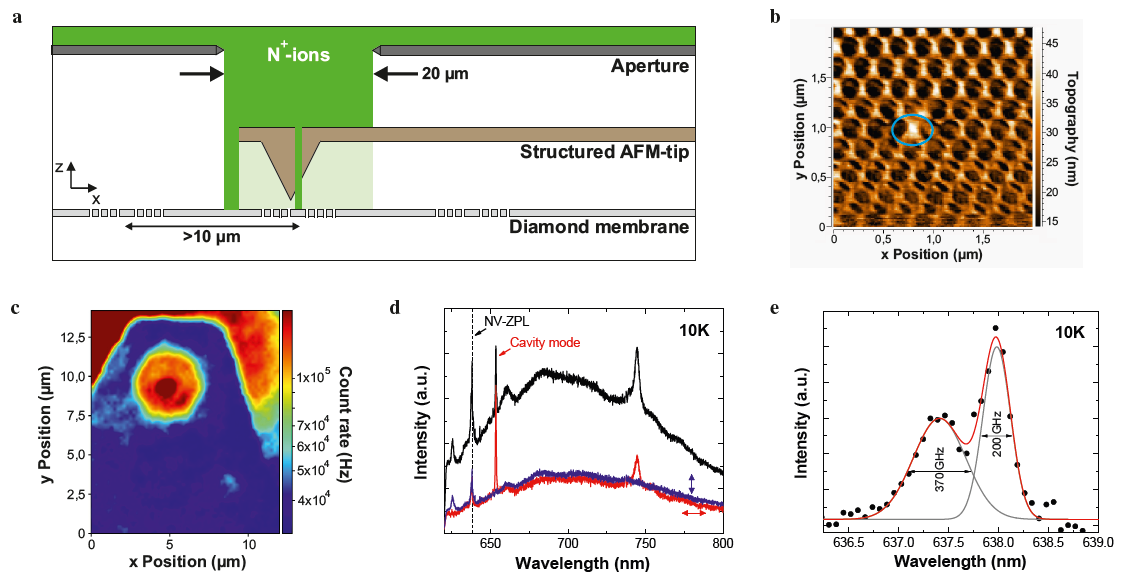}
	\caption{AFM implantation. (a) The pre-focused nitrogen ion beam is at first narrowed down to a diameter of $20\,\upmu\mathrm{m}$ by an aperture and subsequently focused by electrostatic lenses (not sketched) on the backside of the AFM-tip. The final beam diameter of $70\,\mathrm{nm}$ is defined by a FIB-milled hole drilled through the AFM-tip. After the alignment of the AFM-tip relative to the sample, nitrogen ions are implanted into the cavities. The surrounding nanostructure is protected by the AFM-cantilever against unintended ion bombardment. (b) Topography of a M0-cavity recorded by a AFM-scan (non-contact mode). The cavity area is clearly visible (marked blue). (c) Spatially resolved PL image of a M0-cavity with a resolution of $100\,\mathrm{nm}$. The count rate is detected in a filter window between $650\,\mathrm{nm}$ and $750\,\mathrm{nm}$ on the NV sideband and scaled logarithmically. (d and e) PL spectra of the very same M0-cavity at a temperature of $10\,\mathrm{K}$. (d) The cavity mode has a $Q$-factor of $2060$. In addition, a spectral line around $638\,\mathrm{nm}$ is observable, attributed to the NV-ZPL. The blue and the red curve are spectra with a horizontal/vertical polarization filter in the detection path. (e) By zooming into the spectrum around $638\,\mathrm{nm}$ two peaks get clearly visible, which can be fitted by Gaussian lines (gray). The spectra were taken with an integration time of $120\,\mathrm{s}$. The laser excitation in (c-e) was carried out at a wavelength of $532\,\mathrm{nm}$ and a laser power of $500\,\upmu\mathrm{W}$.}
	\label{fig:AFM}
\end{figure*}
As we used ultrapure diamond as starting material, at first nitrogen has to be implanted. In the past few years several high resolution implantation techniques were developed, most of them masking the sample surface by spin-coated photoresist films,\cite{Toyli2010,Spinicelli2011} Mica layers,\cite{Pezzagna2011} transferred silicon hard masks \cite{Bayn2015} as well as structured AFM-tips,\cite{Pezzagna2010a} each featuring small holes as apertures. Also a maskless FIB-implanter was realized.\cite{Lesik2013} The last two techniques have the outstanding advantage, that the implantation spot can be positioned relative to previously produced nanostructures with an accuracy at the nanoscale. Whereas the lateral resolution of a nitrogen FIB implantation is about $100\,\mathrm{nm}$ at a typical acceleration voltage of $30\,\mathrm{kV}$, nitrogen may be implanted through an AFM-tip with a lateral resolution of up to $20\,\mathrm{nm}$ at a typical implantation energy of $5\,\mathrm{keV}$.\cite{Pezzagna2011a} As precise positioning is crucial for efficient cavity coupling, we here use the AFM-tip technique.\\
The home built AFM-setup combines a conventional low-energy ion source for generation and acceleration of nitrogen ions with an AFM-unit. A small hole is FIB-milled close to the apex of the pyramidal AFM-tip which is positioned over the implantation spot (Fig. \ref{fig:AFM}a). To align the AFM-tip relative to the sample, AFM-scans are performed prior to every implantation (Fig.~\ref{fig:AFM}b). The lateral implantation resolution with regard to the diameter of the AFM-aperture of $70\,\mathrm{nm}$, the alignment accuracy of the tip to the sample of about $1\,\mathrm{nm}$, and the ion straggle in diamond of $3\,\mathrm{nm}$ at the chosen implantation energy of $5\,\mathrm{keV}$ yields a total accuracy of $74\,\mathrm{nm}$. Note that small straggling effects of ions at the edges of the $70\,\mathrm{nm}$ aperture are not included in the analysis above and are under present investigation. The expected implantation depth of the nitrogen ions in $(001)$-oriented diamond samples is $13\,\mathrm{nm}$ on average \cite{Lehtinen2016} and the expected yield about $0.8\,\%$.\cite{Pezzagna2010} The implantation dose aiming at one NV center per cavity is calculated to $2.7\cdot 10^{12}\,\mathrm{ions/cm^2}$. As the generation of NV centers is a statistical process, the final implantation doses were varied between half and triple of the calculated value.\\
After implantation, an extensive post-processing is required in order to restore a good diamond crystal quality and to activate NV centers. At first, the samples are therefore annealed in vacuum ($p\leq10^{-6}\,\mathrm{mbar}$) at $900\,^{\circ}\mathrm{C}$ for $10\,\mathrm{h}$. Subsequently, oxidation of the samples at $450\,^{\circ}\mathrm{C}$ for $3\,\mathrm{h}$ in air atmosphere and acid cleaning ($5\,\mathrm{h}$  in a tri-acid mixture of perchloric, sulfuric and nitric acid) removes graphitic residuals from the surface and provides an oxygen termination. Hence, the possibility to obtain negatively charged NV centers within the PhC cavity is enhanced.\cite{Groot-Berning2014}\\
In Fig.~\ref{fig:AFM}c a typical spatially resolved photoluminescence (PL) image after post-processing is shown. Generated NV centers are localized in the PhC cavities as well as outside the area masked by the AFM-cantilever. This observation is in agreement with the results of additionally performed optically detected magnetic resonance (ODMR) measurements, which reveal a dip at frequencies around $2.87\,\mathrm{GHz}$, characteristic for NV centers,\cite{Gruber1997} inside the cavities and no hints of NV centers within the surrounding PhC arrays. The cavities were further examined with PL spectroscopy at low temperatures. In Fig.~\ref{fig:AFM}d and \ref{fig:AFM}e PL spectra of the M0-cavity with the lowest applied implantation dose are shown. Besides the cavity mode spectral lines around $638\,\mathrm{nm}$ are visible featuring Gaussian shapes with half widths (FWHM) of $200\,\mathrm{GHz}$ and $370\,\mathrm{GHz}$, respectively. We performed photon correlation ($g^2$) measurements to estimate the number of generated NV centers. As the $g^2$ results did not show a dip but Poissonian photon statistics we have to assume that a few NV centers were created inside the cavity volume. The deviation from the expected implantation yield might be explained by a large number of defects in the material as a consequence of the PhC fabrication process, providing a high density of vacancies for NV center creation. In addition, the emission lines might be shifted by local strain and strongly broadened by spectral diffusion due to nearby charges.\cite{Fu2010,Robledo2010}

\section{\label{sec:Tuning} Resonance Tuning}

\begin{figure}[!tb]
     \includegraphics[width=6.5cm]{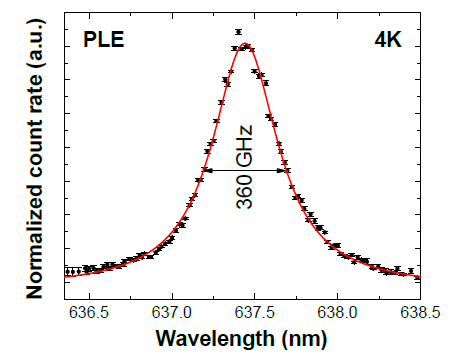}
	\caption{PLE spectrum of the NV-ZPL in the M0-cavity. The count rates detected on the NV sideband in a filter window ranging from $650\,\mathrm{nm}$ to $750\,\mathrm{nm}$ are normalized to the laser power of a tunable diode laser and fitted by a Lorentzian line (red).}
	\label{fig:PLE}
\end{figure}

In the past few years several techniques have been developed for the deterministic tuning of cavity modes into resonance with the ZPL of color centers.
\begin{figure*}[!tb]
\begin{center}
\begin{minipage}[t][9cm][c]{12cm}
     \includegraphics[width=12cm]{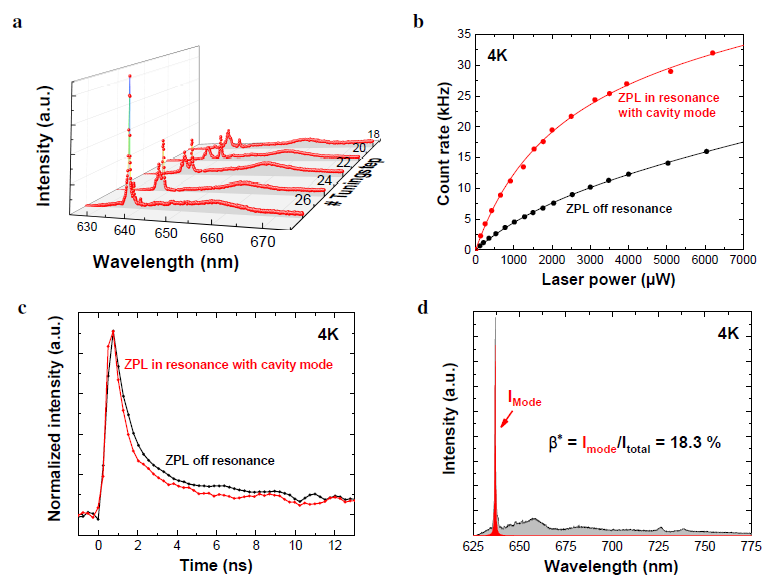}
\end{minipage}
\begin{minipage}[t][9cm][c]{5.5cm}
     	\caption{Tuning the mode of a M0-cavity into resonance with the NV-ZPL at $637.4\,\mathrm{nm}$. (a) PL spectra after different tuning steps. (b) Comparison of a resonant (red) to an off-resonant (black) saturation measurement, (c) comparison of a resonant (red) to an off-resonant (black) lifetime measurement. The off-resonant measurements were performed with the cavity mode tuned to $634\,\mathrm{nm}$. The ZPL photons were detected in a filter window around $638\,\mathrm{nm}$ featuring a spectral width of $2.5\,\mathrm{nm}$. (d) Estimation of  the emission fraction into the cavity mode from the PL spectrum. The ratio of the area below the Lorentzian line (red) and the area below the entire curve (grey+red) yields the corresponding intensity ratio.}
	\label{fig:ErgebnisseTuningC12}
\end{minipage}
\end{center}
\end{figure*}
A well-established but irreversible method is the oxidation of diamond in an air or oxygen atmosphere.\cite{Riedrich-Moeller2012,Hausmann2013} At temperatures above $450\,^{\circ}\mathrm{C}$ the diamond surface starts to oxidize, the thickness of the diamond membrane decreases with an accompanying enlargement of the hole diameters. Larger holes as well as thinner diamond films lead to a spectral blue shift of the cavity modes. Finite Difference Time Domain (FDTD) simulations predict a blue shift of $12\,\mathrm{nm}$ for our M0-cavity if $5\,\mathrm{nm}$ diamond are removed from the surface. On the other hand, cavity modes have been successfully redshifted by condensing inert gases like xenon or nitrogen on the sample's surface,\cite{Lee2014,Hausmann2013,Mosor2005} thereby increasing the sample thickness and reducing the hole radii. In contrast to the oxidation technique, a tuning by gas adsorption is a reversible process where heating the sample enables the residual-free removal of the condensed layers.\cite{Mosor2005}\\
In the following we focus on the M0-cavity implanted with the lowest implantation dose of $1.35\cdot 10^{12}\,\mathrm{ions/cm^2}$, featuring a cavity mode with a $Q$-factor of $2060$ (spectra in Fig.~\ref{fig:AFM}d and \ref{fig:AFM}e). At first, the oxidation technique described above, using a temperature of $525\,^{\circ}\mathrm{C}$, is applied to tune the cavity mode to shorter wavelengths. After tuning and acid-cleaning the resulting mode position is $634\,\mathrm{nm}$. In a second step, the residual gas present in the sample chamber of the cryostat (Attodry 2100, Attocube) is adsorbed to the sample, allowing a redshift of the cavity mode under continuous optical control: we observed that during PL measurements at room-temperature and a pressure of $10^{-4}\,\mathrm{mbar}$ cavity modes show a red shift under continuous laser excitation at $532\,\mathrm{nm}$. Furthermore, only the modes of a cavity directly irradiated with laser light are affected and the observed shift rates scale with the power. All these observations may be explained as follows: as the vacuum pump is attached at the topside of the cryostat, pumping leads to an efficient removal of lightweight molecules, whereas heavy gas molecules, e.\,g. hydrocarbons, partly remain in the sample chamber. Based on the observations we assume a light-assisted adsorption of residual gas molecules onto the sample's surface.\cite{Mosor2005,Preclikova2010} Under illumination with $1.5\,\mathrm{mW}$ of laser light at $532\,\mathrm{nm}$ we observe a shift rate of the cavity modes of approximately $1.8\,\mathrm{nm/h}$. The $Q$-factor remains almost unchanged with $Q=2021$ after tuning. Importantly, after further pumping and cooling to liquid helium temperatures no further mode shifts are observable under laser illumination. We point out that the described tuning by gas adsorption is a reversible process. Heating the sample to $400\,^{\circ}\mathrm{C}$ in air atmosphere resets the sample to the initial state before the gas adsorption.\\
The PL spectra depicted in Fig.~\ref{fig:ErgebnisseTuningC12}a show a strong enhancement of the ZPL intensity by tuning the cavity mode on resonance. We further observe an enhancement of the saturation count rate filtered in a $2.5\,\mathrm{nm}$ wide window by a factor of $2.8$ from $13.6\,\mathrm{kHz}$ to $37.5\,\mathrm{kHz}$ (Fig.~\ref{fig:ErgebnisseTuningC12}b). Furthermore, the emitter's lifetime decreases from $9.0\,\mathrm{ns}$ at a spectral mode position of $634\,\mathrm{nm}$ to $8.0\,\mathrm{ns}$ in resonance with the NV-ZPL. The recorded lifetime traces (Fig.~\ref{fig:ErgebnisseTuningC12}c) feature a double exponential decay. The second time constant on the order of $1\,\mathrm{ns}$ may be attributed to fast decaying fluorescence in the diamond membrane (background). Moreover, PL spectra allow for the estimation of the fraction of emission into the cavity mode as well as the resulting Purcell-factor. The deduced spontaneous emission coupling factor on resonance is $\beta^*=I_{\mathrm{mode}}/I_{\mathrm{total}}=18.3\,\%$ (Fig.~\ref{fig:ErgebnisseTuningC12}d). As the incorporated NV centers feature an off-resonant Debye-Waller factor of $2.1\,\%$, we conclude that the emission fraction into the ZPL is strongly enhanced due to the cavity coupling.\\
The $\beta^*$-factor finally enables us to calculate the total Purcell-factor to be $1+F^*=1.224$. We modeled the total Purcell-factor $1+F^*$ on resonance by using a phonon assisted cavity coupling model presented in \cite{Albrecht2013}. With the $Q$-factor and modal volume $V$ of the cavity mode as well as the measured linewidths of the ZPL and the sideband transitions of the NV centers a Purcell-factor of $1+F^*=4.9$ is predicted. The experimentally determined value of $1+F^*=1.224$ is reduced due to a non-perfect lateral positioning of the implanted NV centers relative to the mode field, a shallow implantation depth as well as a non-perfect dipole orientation in the used (001)-oriented diamond samples. The model further correctly predicts the experimentally observed lifetime reduction.\\
To analyze the NV-ZPL in detail despite the existing resonance with the cavity mode, a photoluminescence excitation (PLE) spectrum was recorded for wavelengths between $636.2\,\mathrm{nm}$ and $638.5\,\mathrm{nm}$. In the spectrum depicted in Fig.~\ref{fig:PLE} it is noticeable, that only one of the previously detected two emission lines (see Fig.~\ref{fig:AFM}e) remains. In the PLE spectrum this line at $637.4\,\mathrm{nm}$ has a half width of $360\,\mathrm{GHz}$, in good agreement with the value of $370\,\mathrm{GHz}$ deduced from the PL spectra. The absence of the second line is most likely due to the removal of shallow implanted NV centers during the extended tuning steps by oxidation.

\section{\label{sec:SNR} SNR enhancement}

To estimate the SNR enhancement due to the emitter-cavity-coupling we at first have to consider the change in the photon count rate $N_0$. For this, we have to take into account the shortening of the emitter's lifetime as well as the change in collection efficiency. Furthermore, for the SNR the change in spin-dependent fluorescence contrast $C$ has to be considered. To this end, a reliable methodology is required to compare the contrast before and after tuning the cavity mode in resonance with the NV-ZPL.

\subsection{\label{Auskopplung}Light extraction from a PhC cavity}

\begin{figure}[!tb]
     \includegraphics[width=7.5cm]{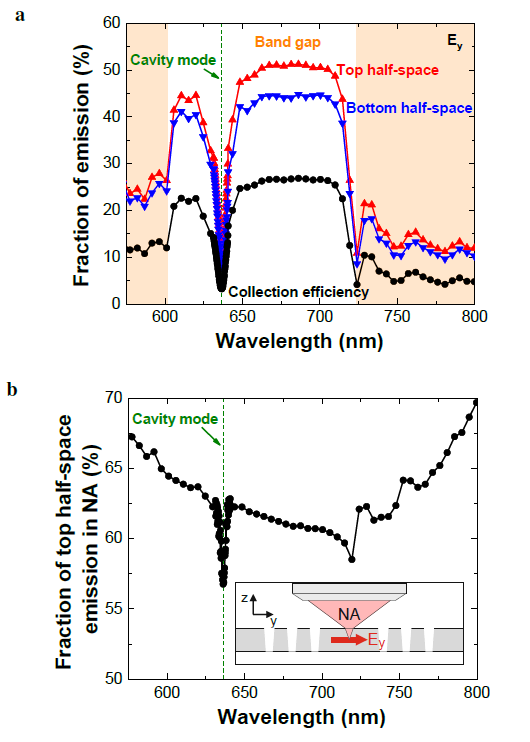}
     	\caption{Simulation of the light emission from a M0-cavity for a dipole oriented in $y$-direction ($E_y$). (a) Collection efficiency (black), fraction of the emitted photons into the top half-space (red) and the bottom half-space (blue). Sectors outside the photonic band gap of the underlying PhC array are marked orange. (b) Fraction of the top half-space light emitted in farfield angles $\leq51.13^{\circ}$, which may be collected by a microscope objective with $\mathrm{NA}=0.8$.}
	\label{fig:SIM}
\end{figure}

We simulate the emission of the cavity-coupled NV centers using a finite element software (FDTD solutions, Lumerical) by modeling the emission of an electric dipole point source positioned in the mode field maximum. The simulated photonic nanostructure features the same number of holes as the produced M0-cavity where the holes have a conical shape with an inclination angle of $4^{\circ}$ (see section \ref{sec:Nanofab}). We calculate the emission fractions in the two half-spaces above and below the PhC as well as the collection efficiency for a microscope objective ($\mathrm{NA}=0.8$) positioned vertically above the diamond membrane. Each simulation is performed for a $E_z$ dipole oriented vertically to the diamond film as well as for the in-plane dipoles $E_x$ and $E_y$. These dipoles are oriented such that the cavity mode is fed by the $E_y$ dipole whereas the $E_x$ dipole has orthogonal polarization. The orthogonality of the dipoles also allows us to calculate the emission for arbitrary dipole orientations as discussed below.\\
As an example we show in Fig.~\ref{fig:SIM}a a simulation for the $E_y$ dipole. We find that the collection efficiency drops to $3.4\,\%$ at the resonance wavelength of the cavity mode, whereas for other wavelengths within the photonic band gap collection efficiencies around $25\,\%$ are achieved.
This minimum arises from a modified directional characteristic of the emitted light. Two factors play a role here: first, the fraction of photons emitted in the top half-space decreases at the mode's resonance wavelength from values over $50\,\%$ to values below $12\,\%$ (Fig.~\ref{fig:SIM}a). Second, a smaller part of the photons emitted in the top half-space (below $57\,\%$ instead of over $62\,\%$) can be collected within the NA at the resonance wavelength of the mode (Fig.~\ref{fig:SIM}b).\\
The NV emission dipoles are oriented in the $(111)$-plane of the diamond lattice. For the small NV ensemble in our experiment we average over all possible dipole orientations within the $(111)$-plane. As the used diamond material features a $(001)$-orientation, the $\langle 111\rangle$-axis exhibits an angle of $35.3^{\circ}$ to the $xy$-plane (membrane plane). The projection of the averaged NV dipoles onto the $x$ or $y$-axis amounts to $24\,\%$ and $52\,\%$ onto the $z$-axis. To calculate the resulting collection efficiency for our NV ensemble, also the Purcell-factors for the three dipole orientations have to be considered. The emission directivity of photons emitted into the cavity mode is only determined by the mode emission directivity. Further, the emission directivity of off-resonant photons is defined by the photonic nanostructure. Considering this, we first calculated the spontaneous emission rates $\tilde{\gamma}_i$ of the projections of the averaged NV dipoles onto the $x$, $y$ and $z$-axis considering the corresponding Purcell-factors $F_i$ as follows:
\begin{eqnarray}
			\tilde{\gamma}_i=\gamma\cdot F_i\cdot k_i\
\end{eqnarray}
with the free-space spontaneous emission rate $\gamma$, $k_i=0.24$ for $i=x,y$ and $k_i=0.52$ for $i=z$.
\begin{figure*}[!tb]
     \includegraphics[width=16cm]{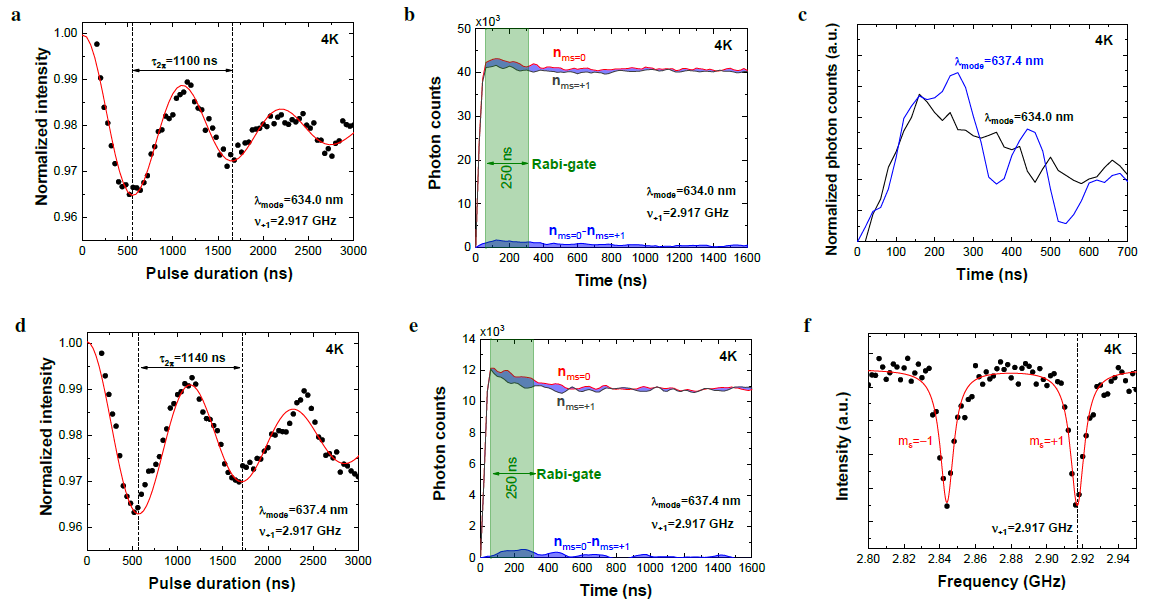}
	\caption{Fluorescence measurements at liquid helium temperature for the cavity-coupled NV centers off (a,b) and on (d,e) resonance. (a and d) Rabi measurements. The NV ensemble is initialized in $m_s=0$ using a $2\,\upmu\mathrm{s}$ laser pulse at $532\,\mathrm{nm}$ followed by a resonant microwave pulse of variable length. During a read-out laser pulse the fluorescence is detected in a $250\,\mathrm{ns}$ wide photon detection gate. The data are fitted by exponentially damped sinusoids (red). (b and e) Time-resolved fluorescence for the NV centers prepared in the $m_s=0$ state (red) and $m_s=+1$ state (black) as well as the resulting time-resolved fluorescence contrasts (blue). The fluorescence contrasts plotted in (c) are normalized to the corresponding photon counts in the steady state. (f) Continuous wave ODMR measurement at an external magnetic field of $2\,\mathrm{mT}$. The excitation is carried out by an off-resonant laser at $532\,\mathrm{nm}$ in the PL saturation regime. The photons are detected on the NV sideband between $650\,\mathrm{nm}$ and $750\,\mathrm{nm}$. The detected photon count rates are background corrected.}
	\label{fig:Kontrast}
\end{figure*}
Subsequently, the fraction of emission produced by the averaged NV dipole projected to the $x$, $y$ or $z$-axis may be calculated by normalizing the spontaneous emission rates $\tilde{\gamma}_i$ to the sum over all these rates. To finally calculate the collection efficiency at the given dipole orientation, the simulated collection efficiencies for a dipole oriented in $x$, $y$ or $z$-direction are weighted with the respective fraction of emission and the resulting values summed up.\\
For the optical read-out of the NV electron spin we can distinguish between two fundamental cases. In the first case only ZPL photons are considered. This is for instance the case, if indistinguishable photons are required. On the other hand, for projective spin read-out generally all the emitted photons are used, e.\,g. if NV centers are used as magnetic field sensors on the nanoscale.\cite{Balasubramanian2008,Fuchs2008,Maze2008,Bernardi2017} The simulations predict a collection efficiency of $3.9\,\%$ for the off-resonant mode position at $634.0\,\mathrm{nm}$, which is slightly reduced to $3.4\,\%$ in resonance with the NV-ZPL. Hence if for the spin read-out only ZPL photons are used, the collection efficiency drops by a factor of $0.87$. If in the second case all emitted photons are used for the spin read-out, the collection efficiency drops only by a factor of $0.97$. The reduction in the second case is smaller, as only a fraction of the emitted photons is affected by the dominant decrease of collection efficiency at the resonance wavelength of the cavity mode.

\subsection{\label{Measurements}Fluorescence contrast measurements}

We apply a protocol composed of three succeeding measurements for the reliable and reproducible determination of the time-resolved, spin-dependent fluorescence contrast of the cavity-coupled NV centers. For the considered off-resonant spectral mode position at $634\,\mathrm{nm}$ at first a ODMR-measurement under continuous laser and microwave excitation is performed. An external magnetic field of $2\,\mathrm{mT}$ is applied to split the two ground state sublevels $m_s=-1$ and $m_s=+1$. Hence, in the ODMR-spectrum in Fig.~\ref{fig:Kontrast}f, two dips are visible belonging to the resonance frequencies of the ground state transitions from $m_s=0$ to $m_s=-1$ and $m_s=+1$ respectively. In the following we only focus on the transition to the $m_s=+1$ sublevel featuring a resonance frequency of $\nu_{+1}=2.917\,\mathrm{GHz}$. For all presented measurements the optical excitation of the NV centers coupled to the M0-cavity is carried out in the PL saturation regime.\\
To determine the population inversion time a Rabi measurement on the transition $m_s=0$ to $m_s=+1$ is performed (Fig.~\ref{fig:Kontrast}a). From the observed damped Rabi oscillations a $\pi$-time of $550\,\mathrm{ns}$ and a spin coherence time of $T_2^*=1.5\,\upmu\mathrm{s}$ is deduced. The value of $T_2^*$ is in accordance with typical spin coherence times observed for shallow NV centers in nanophotonic structures based on single crystal diamond.\cite{Appel2015} This indicates that our PhC fabrication process does not adversely influence the NV spin coherence properties.\\
In Fig.~\ref{fig:Kontrast}b we present the time-resolved, spin-dependent fluorescence. Here, after spin initialization a resonant microwave $\pi$-pulse is applied, resulting in a preparation of the NV centers in the $m_s=+1$ state. During the spin read-out a $50\,\mathrm{ns}$ wide photon detection gate is temporally shifted. For each gate position the measurement is repeated for a NV preparation in the $m_s=0$ state and the difference of the two fluorescence curves, the time-resolved fluorescence contrast, calculated. As expected, the fluorescence curves at first rise for both spin states whereas the count rate for the $m_s=0$ state (\textit{bright state}) is generally higher as for the $m_s=+1$ state (\textit{dark state}) with a maximum contrast of $C=4.2\,\%$. With progressing laser excitation time  both fluorescence curves converge towards a joint value, indicating an equilibrium, spin-mixed state. Whereas the fluorescence contrast measured on single NV centers under zero magnetic field can reach values of $20\,\%$ \cite{Rondin2014} to $30\,\%$ \cite{Dreau2011}, for NV ensembles in a magnetic field contrasts of only a few percent are expected.\cite{Chipaux2015} Possible reasons are lattice and surface defects introduced by FIB milling, impairing charge state stability and spin coherence of the shallow implanted emitters.\cite{Ofori-Okai2012,Chipaux2015} Furthermore, a strain gradient may locally alter the ground state splitting. Hence, different emitters of the considered small NV ensemble may feature slightly different resonance frequencies.\\
Subsequently, the cavity mode is tuned into resonance with the NV-ZPL (see section \ref{sec:Tuning}). As for the tuning process the sample has to be dismounted from the cryostat, a repositioning of the microwave antenna, an air suspended gold wire loop mounted on a positioner, is required after remounting the sample. To establish comparable experimental conditions we position the antenna in such a way  that a $\pi$-time of $570\,\mathrm{ns}$ is reached (Fig.~\ref{fig:Kontrast}d), in good agreement with the $\pi$-time of $550\,\mathrm{ns}$ (Fig.~\ref{fig:Kontrast}a) for the NV-ZPL off resonance with the cavity mode. We follow the same protocol as in the off-resonant case to measure Rabi oscillations (Fig.~\ref{fig:Kontrast}d) as well as fluorescence curves (Fig.~\ref{fig:Kontrast}e) on resonance. When comparing the resulting time-resolved fluorescence contrast on- and off-resonance (Fig.~\ref{fig:Kontrast}c), we find a small reduction of contrast of $1.5\,\%$ on resonance (integrated over the entire time interval in Fig.~\ref{fig:Kontrast}c).

\subsection{\label{Modelling}Rate equation model}

\begin{figure*}[!tb]
\begin{center}
\begin{minipage}[t][9cm][c]{12cm}
     \includegraphics[width=12cm]{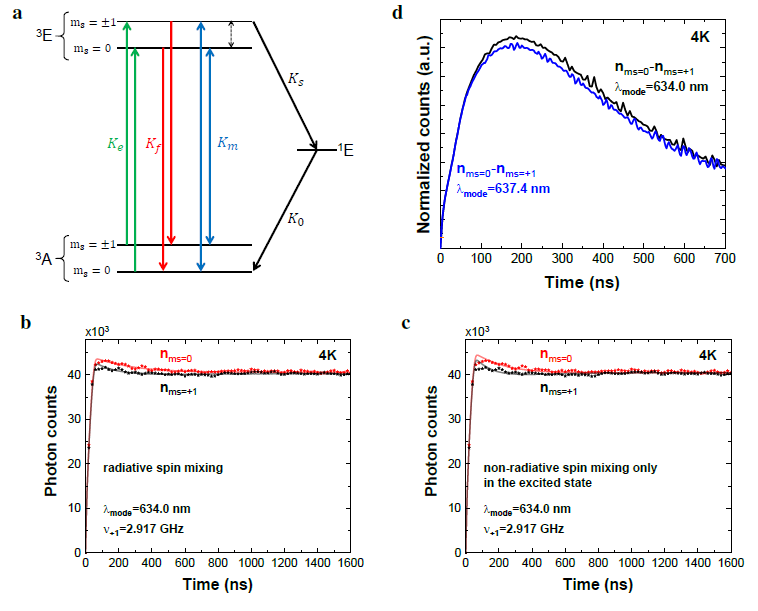}
\end{minipage}
\begin{minipage}[t][9cm][c]{5.5cm}
     	\caption{Rate equation model. (a) NV center modeled as a five level system. The radiative triplet transitions (both spin-conserving (green, red) and spin-mixing (blue)) are illustrated as colored and the ISC transitions as black arrows. The dashed arrow indicates a scenario where non-radiative spin-mixing happens in the excited state. (b and c) Measured time-resolved, spin-dependent fluorescence after preparing the NV centers in the $m_s=0$ state (red dots) and $m_s=\pm1$ state (black dots) for the NV centers off resonance with the cavity mode. The modeled curves (red and black lines) were adapted for the rates $K_e=K_f=111\,\mathrm{MHz}$ under the assumption of a radiative spin-mixing (b) and a non-radiative spin-mixing in the excited state (c), respectively. (d) Simulated fluorescence contrast of NV centers off (black curve) and on (blue curve) resonance with the cavity mode under the assumption of a spin-mixing due to ISC and by radiative transitions in the triplet system.}
	\label{fig:KontrastModellWolf}
\end{minipage}
\end{center}
\end{figure*}

To analyze the measured spin-dependent fluorescence contrast we set up a rate equation model following Wolf et al. .\cite{Wolf2015} In this model a NV center is adopted as a five level system (Fig.~\ref{fig:KontrastModellWolf}a), consisting of the two ground states $m_s=0$ and $m_s=\pm1$, respectively, the corresponding excited states and a long-living singlet state. For the modeling we assume that the excitation rates out of the two ground states are identical for the non-resonant laser excitation.\cite{Manson2006} Furthermore it is assumed that the emission rates from the excited to the ground states of the triplet system are spin independent.\cite{Goldman2015} In addition, the transition rate from the excited $m_s=0$ state to the singlet state is neglected as this rate is four orders of magnitude smaller than the corresponding transition rate for the $m_s=\pm1$ state.\cite{Young2009,Goldman2015,Kalb2018} As a further assumption the transition rate from the singlet state to the $m_s=\pm1$ ground state is neglected, because the rate to the $m_s=0$ ground state is about six times larger.\cite{Thiering2018,Kalb2018} By reason of these assumptions a spin-mixing, leading to a statistical mixture of $m_s=0$ and $m_s=\pm1$ states due to the spin-selective intersystem crossing is already included in the model. Such a spin-mixing was both theoretically predicted \cite{Wolf2015,Bogdanov2017,Thiering2018} and experimentally confirmed.\cite{Kalb2018}\\
If, however, the only spin-mixing process was due to ISC, the modeled population would end up in the $m_s=0$ state for sufficient long-lasting laser excitation and the modeled fluorescence would thus be maximized. Instead, the spin-dependent fluorescence curves in our experiment (Fig.~\ref{fig:Kontrast}b and \ref{fig:Kontrast}e) as well as for other NV centers (see e.\,g. \cite{Steiner2010,Robledo2011a}) show a convergence of the fluorescence and hence the populations towards a steady state. Therefore a further spin-mixing has to be included in the model. Wolf et al. propose as additional spin-mixing mechanisms on the one hand a radiative spin-mixing between the excited and ground states of the triplet system and on the other hand a purely non-radiative spin-mixing between the excited states.\cite{Wolf2015} As can be seen in Fig.~\ref{fig:KontrastModellWolf}b and \ref{fig:KontrastModellWolf}c, respectively, the measured data may be fitted well under the assumption of a radiative spin-mixing and, in particular, better than under the assumption of a non-radiative spin-mixing in the excited state. However, Kalb et al. demonstrated that a radiative spin-mixing occurs only with a probability of well below $1\,\%$ and hence should not be the dominant spin-mixing mechanism in our experiments.\cite{Kalb2018} Instead, the reason for the additionally observed spin-mixing mechanism is that in our experiments we apply an external magnetic field to split the ground states of the NV centers (see section \ref{Measurements}). As the diamond sample features a $(001)$-surface and the magnetic field an orientation vertical to the diamond's surface, the magnetic field exhibits an angle of here $54.7^{\circ}$ to the magnetic dipole axis of the NV centers. As now both ground and excited states consist of superpositions of bare, zero-field spin states, optical transitions with spin flips become allowed resulting in an effective radiative spin mixing.\cite{Tetienne2012} Furthermore, the rates of ISC transitions are modified. Therefore, in summary, the data are fitted appropriately under the assumption of a spin-mixing due to ISC and by radiative transitions in the triplet system.\\
In the following we assume for simplicity that the spin-mixing induced by the off-axis magnetic field is the same for excitation and emission.
With this, the following rate equations are set up, determining the internal dynamics of a NV center:
\begin{eqnarray}
	\dot{B}_{G,0} & = & -(K_e+2K_m)B_{G,0}+K_fB_{E,0}+K_mB_{E,1}+K_0B_S \nonumber \\
	\dot{B}_{G,1} & = & -(K_e+K_m)B_{G,1}+K_fB_{E,1}+2K_mB_{E,0} \nonumber \\
	\dot{B}_{E,0} & = & K_eB_{G,0}-(K_f+2K_m)B_{E,0}+K_mB_{G,1}  \\
	\dot{B}_{E,1} & = & K_eB_{G,1}-(K_f+K_s+K_m)B_{E,1}+2K_mB_{G,0} \nonumber \\
	\dot{B}_{S} & = & K_sB_{E,1}-K_0B_S \nonumber \ .
\end{eqnarray}
$B_{G,0}$ and $B_{G,1}$ are the populations of the ground states $m_s=0$ and $m_s=\pm1$, $B_{E,0}$ and $B_{E,1}$ the populations of the corresponding excited states and $B_S$ the population of the singlet state. $K_e$ and $K_f$ are the spin-preserving excitation and emission rates, $K_s$ the transition rate from the excited $m_s=\pm1$-state to the singlet state, $K_0$ the transition rate from the singlet state to the $m_s=0$ ground state and $K_m$ the rate of the radiative spin-mixing induced by the off-axis magnetic field. As the measurements are performed in the PL saturation regime, the relation $K_e=K_f$ holds in the following.\\
According to the measured off-resonant lifetime value of $9.0\,\mathrm{ns}$, a decay rate of $K_f=111\,\mathrm{MHz}$ is set in the model. With the parameters $K_e$ and $K_f$ fixed, the rate equations are solved and the simulated time-resolved fluorescence is fitted to the measured curves by varying the free parameters $K_0$, $K_s$ and $K_m$ (Fig.~\ref{fig:KontrastModellWolf}b). From this fit a spin-mixing rate of $K_m=1.35\,\mathrm{MHz}$, a transition rate to the singlet state of $K_s=1.79\,\mathrm{MHz}$ and a transition rate to the $m_s=0$ ground state of $K_0=5.80\,\mathrm{MHz}$ are determined.\\
These rates allow us to theoretically predict the time-resolved, spin-dependent fluorescence contrast on resonance. Fig.~\ref{fig:KontrastModellWolf}d shows the fluorescence contrast simulated by the rate equation model for the cavity mode on and off resonance with the NV-ZPL. The theoretically predicted fluorescence contrast on resonance is slightly lower than for the off-resonant case. This is expected, as the emission rate $K_f$ scales with the Purcell-factor and the probability for spin-mixing is the larger the more fluorescence cycles occur per time unit. Depending on the position as well as the size of the temporal read-out window, a reduction of the fluorescence contrast by tuning the mode into resonance with the NV-ZPL of up to $5\,\%$ is theoretically predicted. For a realistic read-out photon gate of $250\,\mathrm{ns}$, as for instance typically used for the acquisition of Rabi measurements, the fluorescence contrast would be reduced by $4.1\,\%$. Such a reduction is conformable with the experimental results (see section \ref{Measurements}).

\subsection{\label{SNRenhancement}Implications for the SNR}

We now discuss the modification of the spin measurement SNR induced by the modified photon collection and modified spin-dependent fluorescence contrast due to the cavity coupling.
With equation (\ref{eq:SNRDefinition}) as well as the appropriate assumption of small contrasts, we arrive at:
\begin{eqnarray}
	\zeta=\frac{\mathrm{SNR^*}}{\mathrm{SNR}}\approx \sqrt{\frac{N_0^*}{N_0}}\cdot \frac{C^*}{C}
\end{eqnarray}
where variables with (without) a star indicate the on- (off-) resonant case. Based on the presented measurements as well as the theoretical modeling we assume in the following a reduction of the spin-dependent fluorescence contrast by $4.1\,\%$, i.\,e. $C^*/C=0.959$. Due to the lifetime reduction the number of emitted photons is increased by a factor of $1.13$, when the mode is tuned into resonance with the NV-ZPL. If we, in a first scenario, consider only ZPL photons for spin read-out, the collection efficiency is reduced by a factor of $0.87$ (see section \ref{Auskopplung}). Furthermore, the fraction of photons emitted into the ZPL is modified from $2.1\,\%$ off resonance to $18.3\,\%$ on resonance (see Fig.~\ref{fig:ErgebnisseTuningC12}d), resulting in an enhancement factor of $8.7$. The total estimated enhancement of the collected photons by applying a narrow-band photon detection around the ZPL is hence $N_0^*/N_0=1.13\cdot0.87\cdot8.7=8.5$ and the SNR, also taking the contrast reduction of $C^*/C=0.959$ into account, thus enhanced by a factor of $\zeta=2.8$. As a result, the SNR is almost tripled by tuning the mode into resonance with the NV-ZPL. If we, in a second scenario, consider the collection of all photons for the spin read-out, we cannot benefit from the higher emission fraction into the ZPL any more, but the number of emitted photons is still enhanced by a factor of $1.13$ due to the lifetime shortening on resonance. The collection efficiency is here reduced by a factor of $0.97$ (see section \ref{Auskopplung}). Altogether the number of detected photons by applying a broad-band photon detection is enhanced by a factor of $N_0^*/N_0=1.13\cdot0.97=1.10$, resulting in a very small SNR enhancement of about $0.5\,\%$. In summary, the cavity coupling of NV centers as demonstrated here leads to a spin read-out SNR enhancement of up to a factor of $\approx3$. This factor is within the same order of magnitude as the SNR enhancement achieved by other methods for improving the photon collection efficiency.\cite{Jamali2014,Robledo2011,Steiner2010}

\section{\label{sec:Summary} Conclusions}

In summary, we reported on the SNR enhancement of the optical spin read-out achieved by tuning the mode of a two-dimensional PhC cavity into resonance with the NV-ZPL. To achieve this, ultrapure $(001)$-oriented CVD-grown diamond films were used as starting material for the RIE fabrication of thin, air-suspended membranes. An extended characterization allowed us to select defect-free spots featuring a suitable thickness for the subsequent fabrication of PhC cavities by FIB milling. The analyzed cavity modes showed $Q$-factors of up to $8250$ at mode volumes of less than one cubic wavelength. The application of a high-resolution implantation technique using a pierced AFM-tip allowed the subsequent generation of NV centers in the cavities. The combination of two spectral tuning methods, an oxidation technique for the blue shift and a gas adsorption technique for the red shift, facilitated the reliable and precise tuning of a cavity mode. For the considered cavity-coupled NV centers the SNR was almost tripled. A theoretical model, taking into account the measured shortening of the emitter’s lifetime, the measured and theoretically predicted change in fluorescence contrast as well as the simulated modification of the collection efficiency reproduces the experimental findings very well.\\
Whereas the reported SNR enhancement is on par with simpler methods for photon collection enhancement,\cite{Jamali2014,Steiner2010,Robledo2011} it could be still increased by a further optimized cavity coupling: a $(111)$-oriented diamond sample with an optimal dipole orientation may lead to a fourfold enhancement of the total Purcell-factor. For an optimal implantation depth in the center of the diamond film, instead of the shallow implantation, the Purcell-factor may be further increased by a factor of $3.2$. In total, for an optimal cavity coupling our theoretical model predicts a SNR enhancement by a factor of more than $6$. A possible method to reach this is a change of the fabrication order such that at first NV centers are implanted into the diamond film and subsequently a PhC cavity is fabricated around. This would allow a maskless implantation with higher energies resulting in deeper implanted NV centers. Subsequently, NV centers could be pre-characterized and suitable single emitters chosen, featuring an ideal dipole orientation.  Alternatively, also improved nanoimplantation techniques such as a maskless FIB-implantation of ions with high lateral resolution are in reach.\cite{Schroeder2017a} Eventually, the here generated NV centers feature an optical linewidth of several hundred GHz, whereas narrow line widths below $100\,\mathrm{MHz}$ are accessible,\cite{Ruf2019} at least in $\upmu$m-thin RIE-etched diamond membranes. Smaller linewidths would further increase the total Purcell-factor. Therefore, in conclusion, higher SNR enhancements are within reach with the method presented here.

\begin{acknowledgments}
We thank B. Lägel and S. Wolff (Nano Structuring Center, University of Kaiserslautern) for helpful discussions on nano-fabrication and use of their facilities. We further thank Alexander Huck and Simeon Bogdanov for helpful discussions on fluorescence contrast measurements and modeling. This research has been partially funded by the European Quantum Technology Flagship Horizon 2020 (H2020-EU1.2.3/2014-2020) under Grant No. 820394 (ASTERIQS). EU funding for the project AME-Lab (European Regional Development Fund C/4-EFRE 13/2009/Br) for the FIB/SEM is acknowledged. E. Neu acknowledges funding via the NanoMatFutur program of the German Ministry of Education and Research (BMBF) under Grant No. FKZ13N13547 as well as a PostDoc Fellowship by the Daimler and Benz Foundation.
\end{acknowledgments}

\nocite{*}
\bibliography{aipsamp}

\end{document}